\documentclass[3p,times,procedia]{elsarticle}
\usepackage{nupha_ecrc}

\volume{00}

\firstpage{1}

\journalname{Nuclear Physics A}

\runauth{}

\jid{nupha}

\jnltitlelogo{Nuclear Physics A}

\usepackage{amssymb}
\usepackage{subfigmat}

\usepackage[figuresright]{rotating}

\begin{document}

\begin{frontmatter}



\dochead{}

\title{Overview of recent ALICE results}


\author{Taku Gunji on behalf of the ALICE Collaboration}
\fntext[col1] {A list of members of the ALICE Collaboration and acknowledgements can be found at the end of this issue.}
\address{Center for Nuclear Study, the University of Tokyo, 7-3-1 Hongo, Bunkyo-ku, Tokyo, 113-0033, Japan}

\begin{abstract}
The ALICE experiment explores the properties of strongly interacting QCD matter at extremely high temperatures 
created in Pb-Pb collisions at LHC and provides further insight into small-system physics in (high-multiplicity)
pp and p-Pb collisions. The ALICE collaboration presented 27 parallel talks, 50 posters, and 1 flash talk
at Quark Matter 2015 and covered various topics including collective dynamics, correlations and fluctuations, 
heavy flavors, quarkonia, jets and high $p_{\rm T}$ hadrons, electromagnetic probes, small system physics, and 
the  upgrade program. This paper highlights some of the selected results.
\end{abstract}


\end{frontmatter}


\section{Introduction}
\label{sec:intro}
ALICE (A Large Ion Collider Experiment) is one of the major experiments at the CERN-LHC (Large Hadron Collider). 
ALICE is dedicated to the study of heavy-ion physics. 
The details of the ALICE detectors and performances in Run1 are described in Ref~\cite{bib:bib1, bib:bib2}. 
The central features of the ALICE detectors are efficient charged particle tracking down to very low 
$p_{\rm T} \sim$ 0.15 GeV/$c$, excellent particle identification (hadrons, electrons, muons, and photons) over a wide momentum range, and excellent
vertexing for the measurements of V$^{0}$, cascades, heavy flavors, and conversion electrons from photons. 

The ALICE collaboration has presented new and exciting results at Quark Matter 2015, which extend our knowledge on 
the dynamics of ultra-relativistic proton-proton and heavy-ion collisions. This paper is an overview of  
27 parallel talks, 50 posters, and 1 flash talk delivered by the ALICE collaboration.

\section{Highlights from Pb-Pb collisions}
\subsection{Collective Dynamics and Correlations}

\begin{figure}[htbp]
\begin{subfigmatrix}{3}
\subfigure{\includegraphics[width=0.32\hsize]{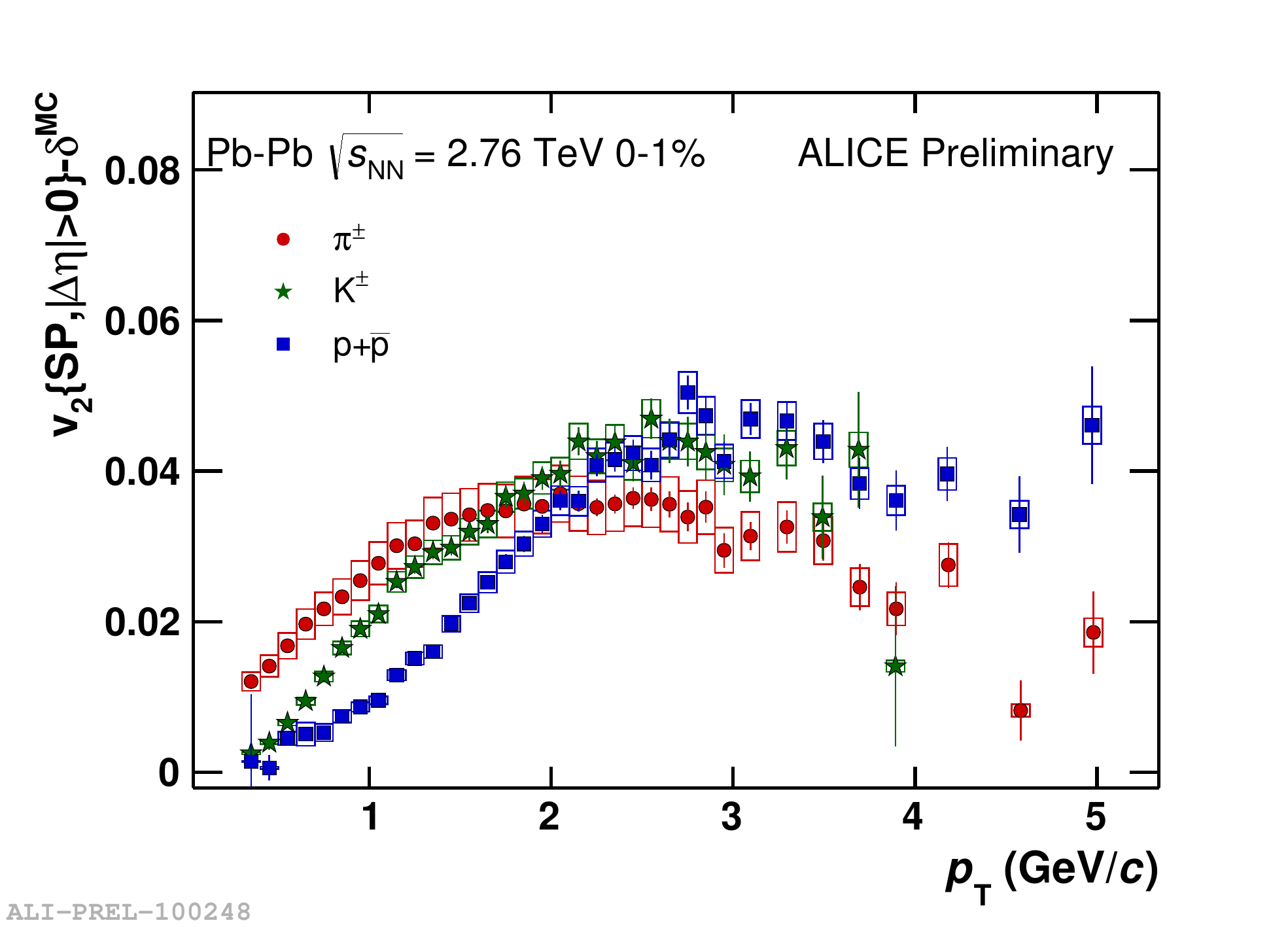}
\label{fig:figa11}}
\subfigure{\includegraphics[width=0.32\hsize]{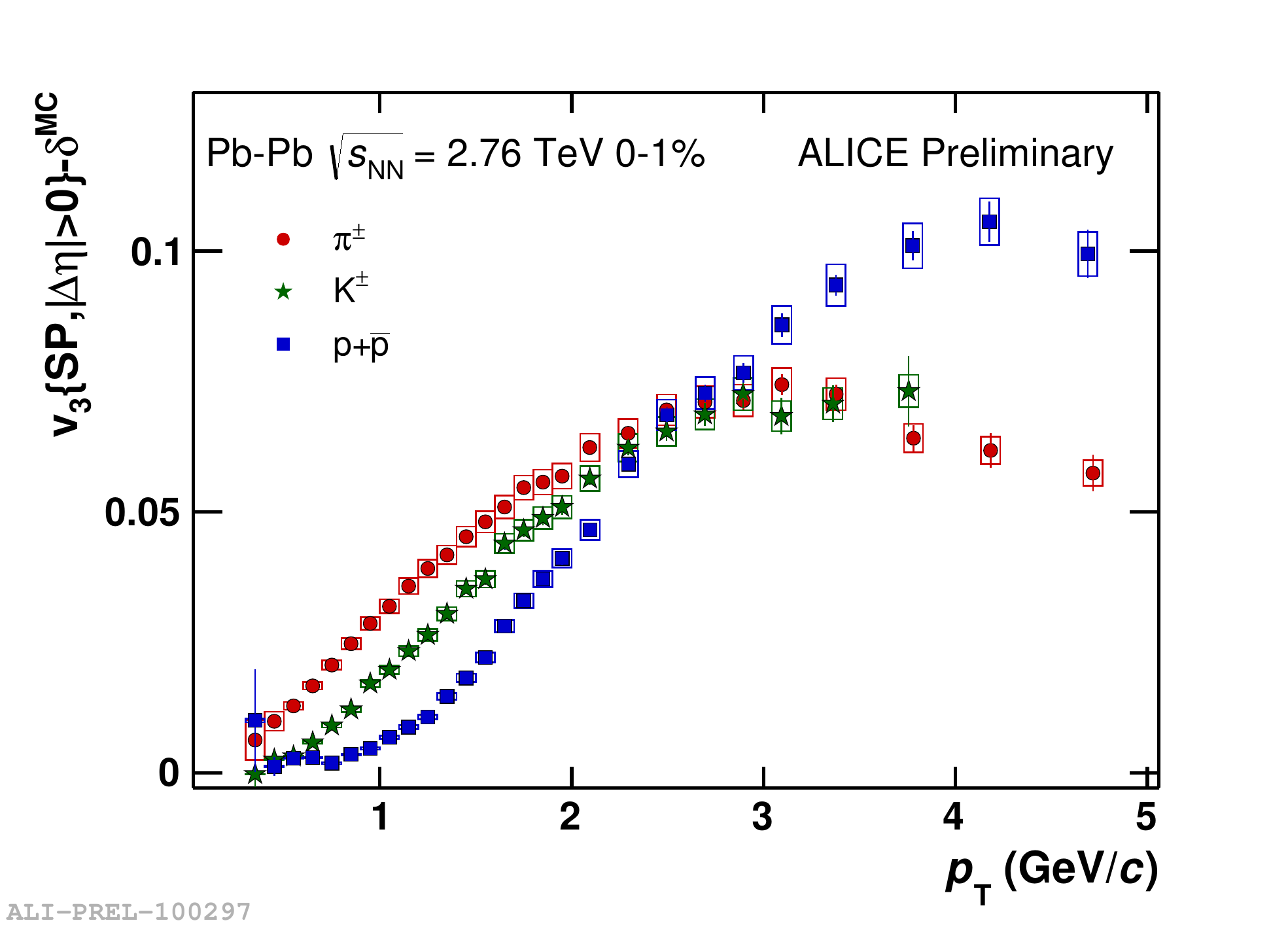}
\label{fig:figa12}}
\subfigure{\includegraphics[width=0.32\hsize]{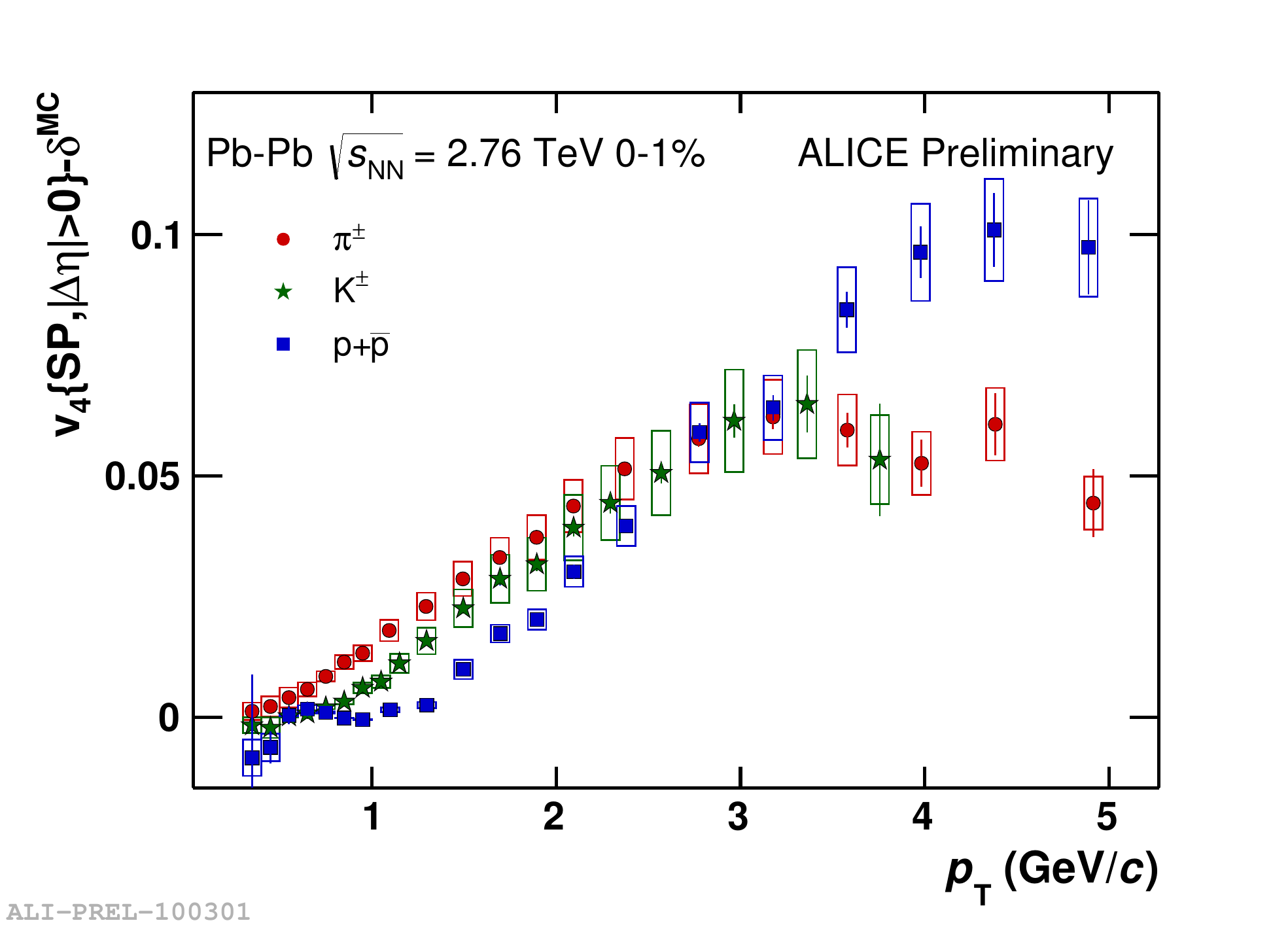}
\label{fig:figa13}}
\end{subfigmatrix}

\begin{subfigmatrix}{3}
\subfigure{\includegraphics[width=0.32\hsize]{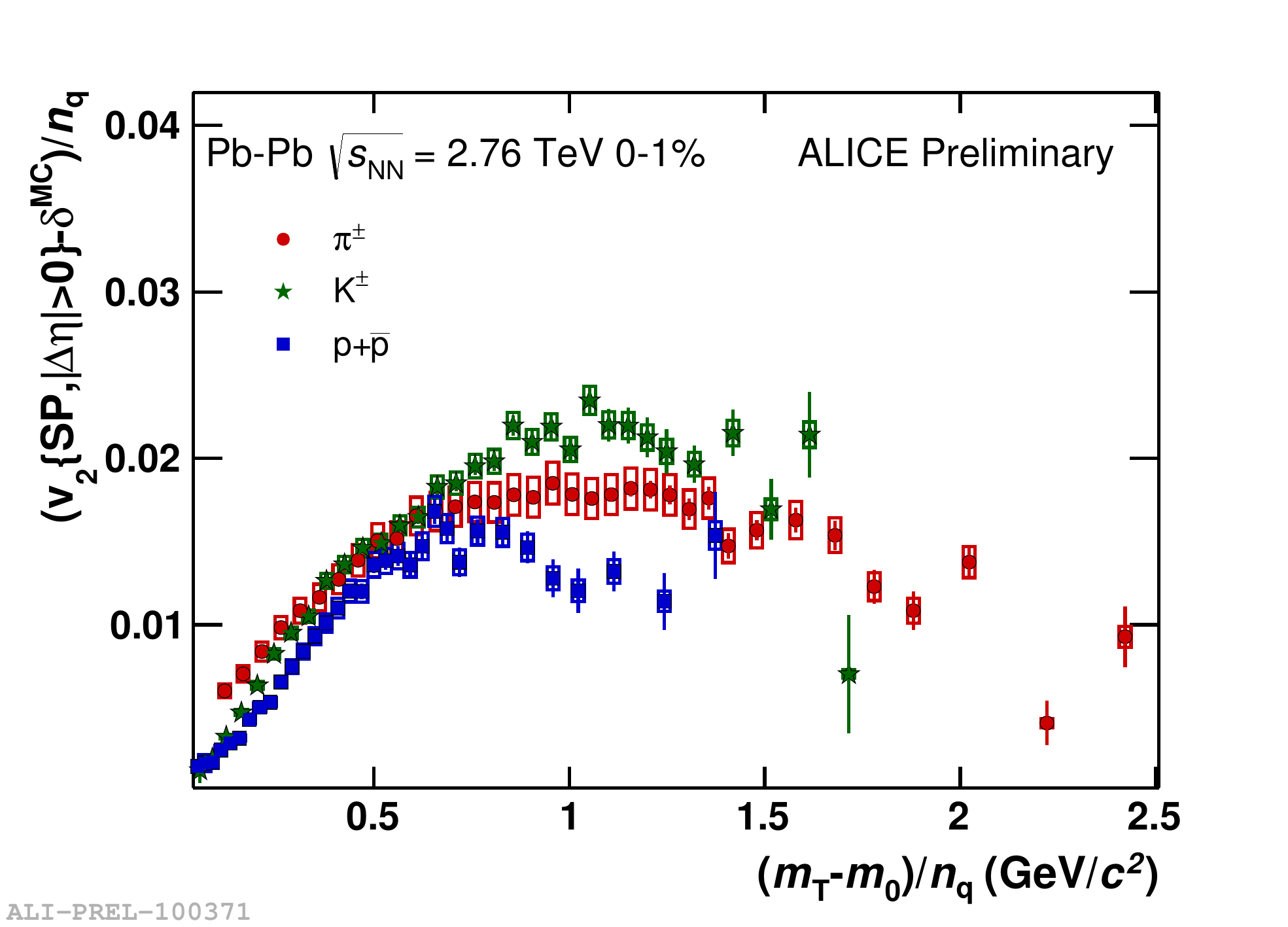}
\label{fig:figa21}}
\subfigure{\includegraphics[width=0.32\hsize]{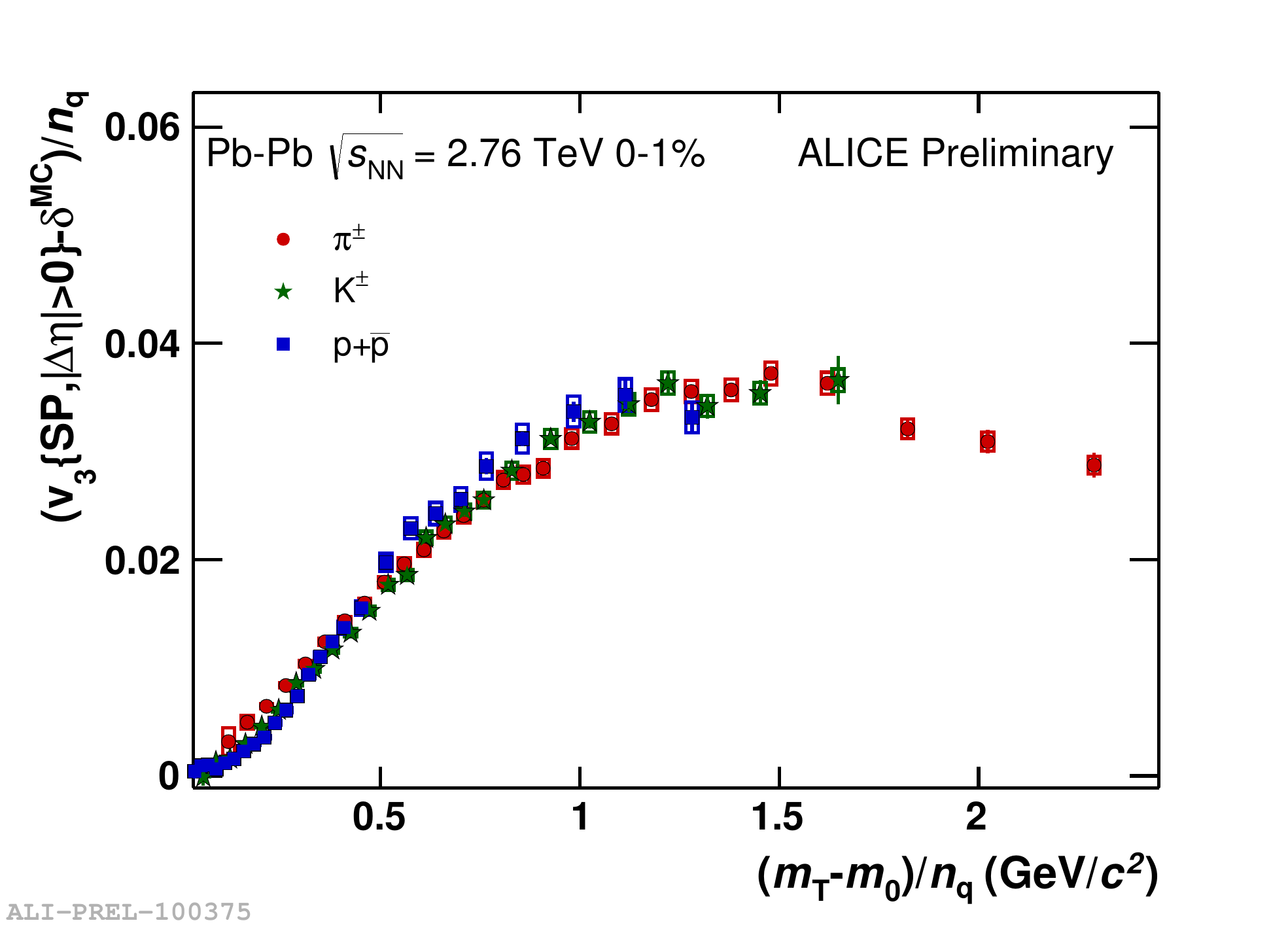}
\label{fig:figa22}}
\subfigure{\includegraphics[width=0.32\hsize]{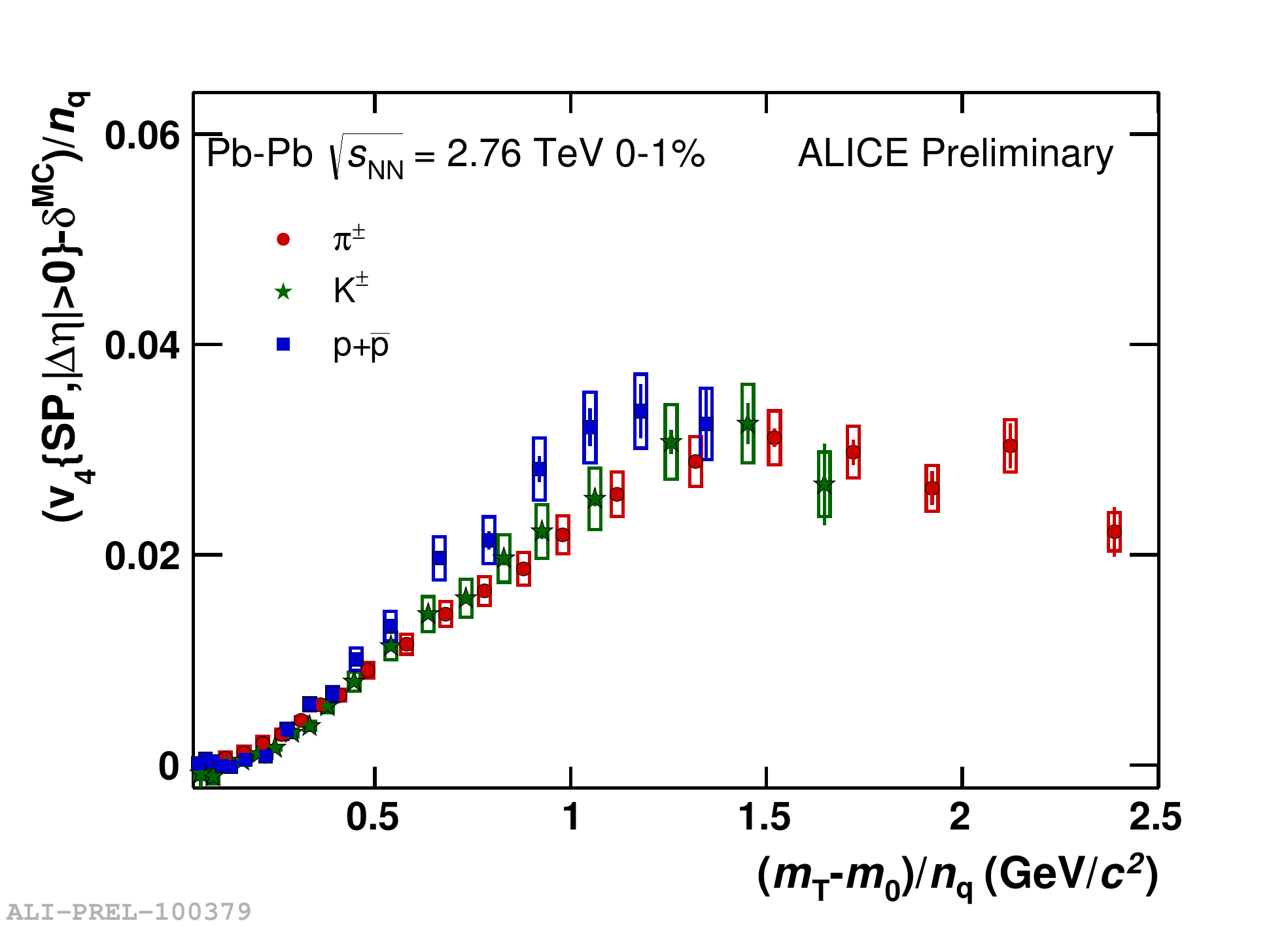}
\label{fig:figa23}}
\end{subfigmatrix}
\caption{(Color online) 
Upper: $v_{2}$ (left), $v_{3}$ (middle), and $v_{4}$ (right) for $\pi$, $K$, and protons in 0-1\% centrality.
Lower: $KE_{\rm T}/nq$ scaling for $v_{2}$ (left), $v_{3}$ (middle), and $v_{4}$ (right) for $\pi$, $K$, and protons in 0-1\% centrality.}
\label{fig:fig2}
\end{figure}

The first results on the $p_{\rm T}$ differential $v_{2}$, $v_{3}$, and $v_{4}$ 
for $\pi^{\pm}$, K$^{\pm}$, and $p+\bar{p}$ for top 0-1\% and 20-30\% centrality classes have been
shown (upper panels of Fig.~\ref{fig:fig2})~\cite{bib:CC1}.
These flow harmonics have been measured with the scalar product method with a pseudo-rapidity gap of $|\Delta \eta| \ge$0 
applied between the identified particles and the reference charged particles. 
The contribution from non-flow effects is estimated using the HIJING event generator and corrected for. 
A clear mass ordering is seen in the low $p_{\rm T}$ region (for $p_{\rm T} \le$ 3 GeV/$c$) for all $v_{n}$.
The lower panels of Fig.~\ref{fig:fig2} shows $KE_{\rm T}/n_{q}$ scaling 
for $v_{2}$ (left), $v_{3}$ (middle), and $v_{4}$ (right) 
for $\pi^{\pm}$, K$^{\pm}$, and $p+\bar{p}$ in 0-1\% centrality. 
One can see that $KE_{\rm T}/n_{q}$ scaling works differently for different $v_{n}$ and works best for $v_{3}$. 


\begin{figure}[htbp]
\begin{subfigmatrix}{2}
\subfigure{\includegraphics[width=0.44\hsize]{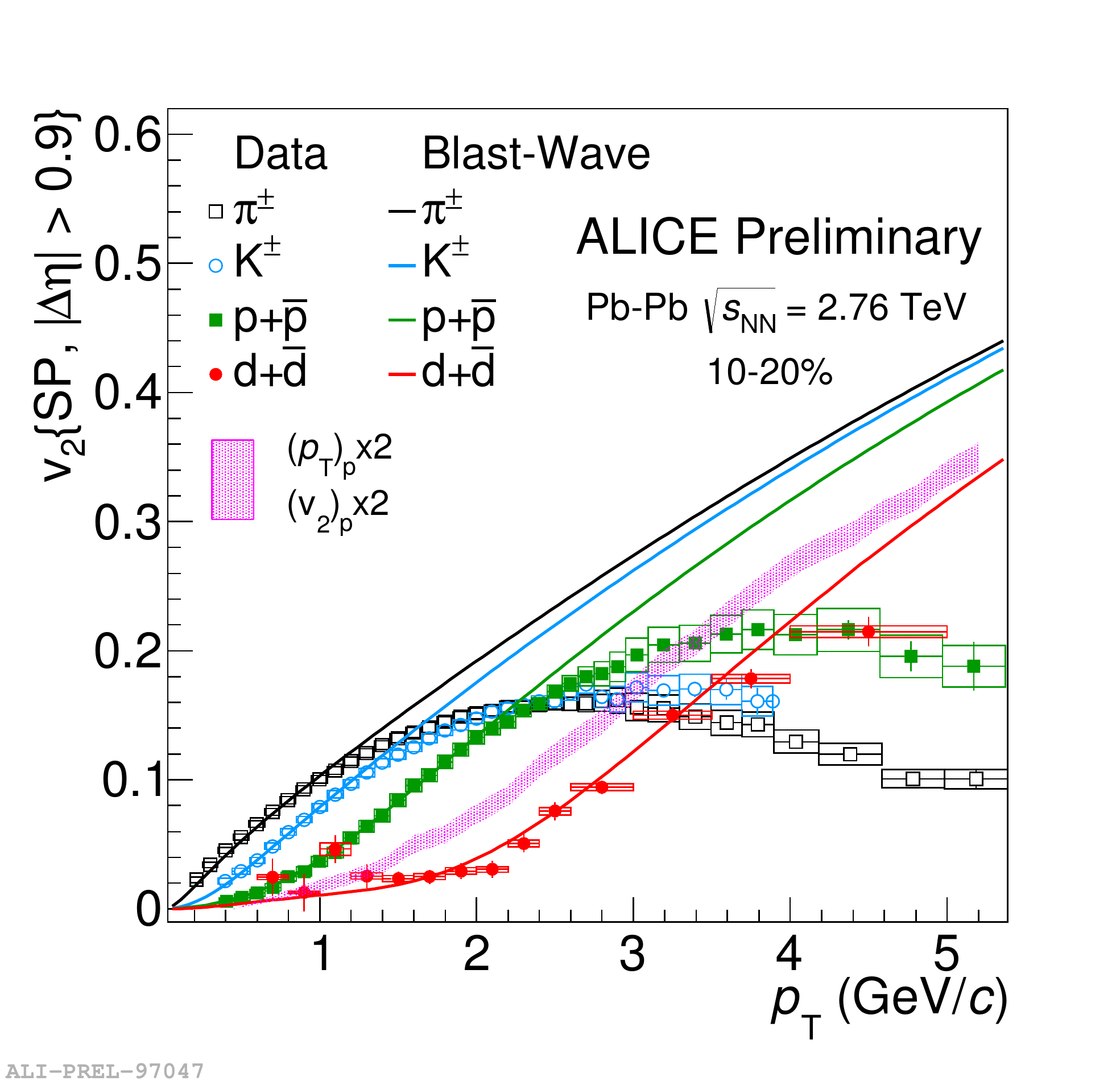}
\label{fig:figa31}}
\subfigure{\includegraphics[width=0.44\hsize]{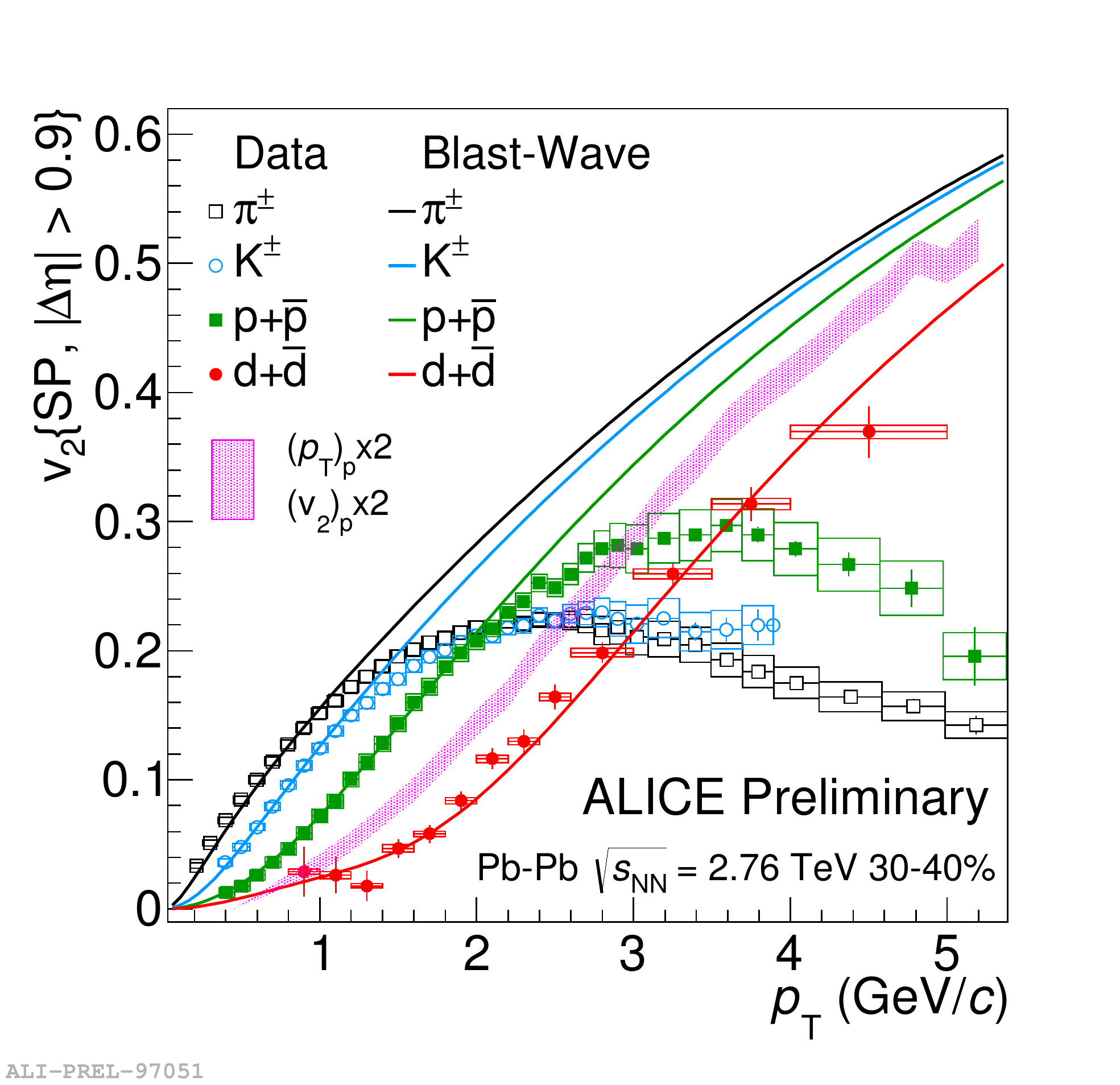}
\label{fig:figa32}}
\end{subfigmatrix}
\caption{(Color online) Deuteron $v_{2}$ as a function of $p_{\rm T}$ and $v_{2}$ expected from simple coalescence model for 10-20 \% (left) and 30-40 \% centrality (right).}
\label{fig:fig3}
\end{figure}

Left and right of Fig.~\ref{fig:fig3} show deuteron (${\rm d}$+$\bar{\rm d}$), proton ($p$+$\bar{p}$), K$^{\pm}$, 
and $\pi^{\pm}$ $v_{2}$ as a function of $p_{\rm T}$ in 10-20\% and 30-40\% centrality classes, 
respectively~\cite{bib:CC2}.
Solid lines are the results of the Blast-Wave model fit to the measured $\pi^{\pm}$, K$^{\pm}$, and proton 
$p_{\rm T}$ spectra and $v_{2}$. The deuteron $v_{2}$ is then calculated using the resulting Blast-Wave 
model fit parameters. 
Hatched area shows the $p_{\rm T}$ differential $v_{2}$ estimated by a simple coalescence model, 
where $v_{2}$ of deuteron is 
calculated as $v_{2} (p_{\rm T})_{\rm deuteron} = 2v_{2}(2 {p_{\rm T}}_{\rm proton})_{\rm proton}$.
Deuteron $v_{2}$ is well-described by the Blast-Wave model and 
is overestimated by the simple coalescence model.

\begin{figure}[htbp]
\begin{subfigmatrix}{2}
\subfigure{\includegraphics[width=0.485\hsize]{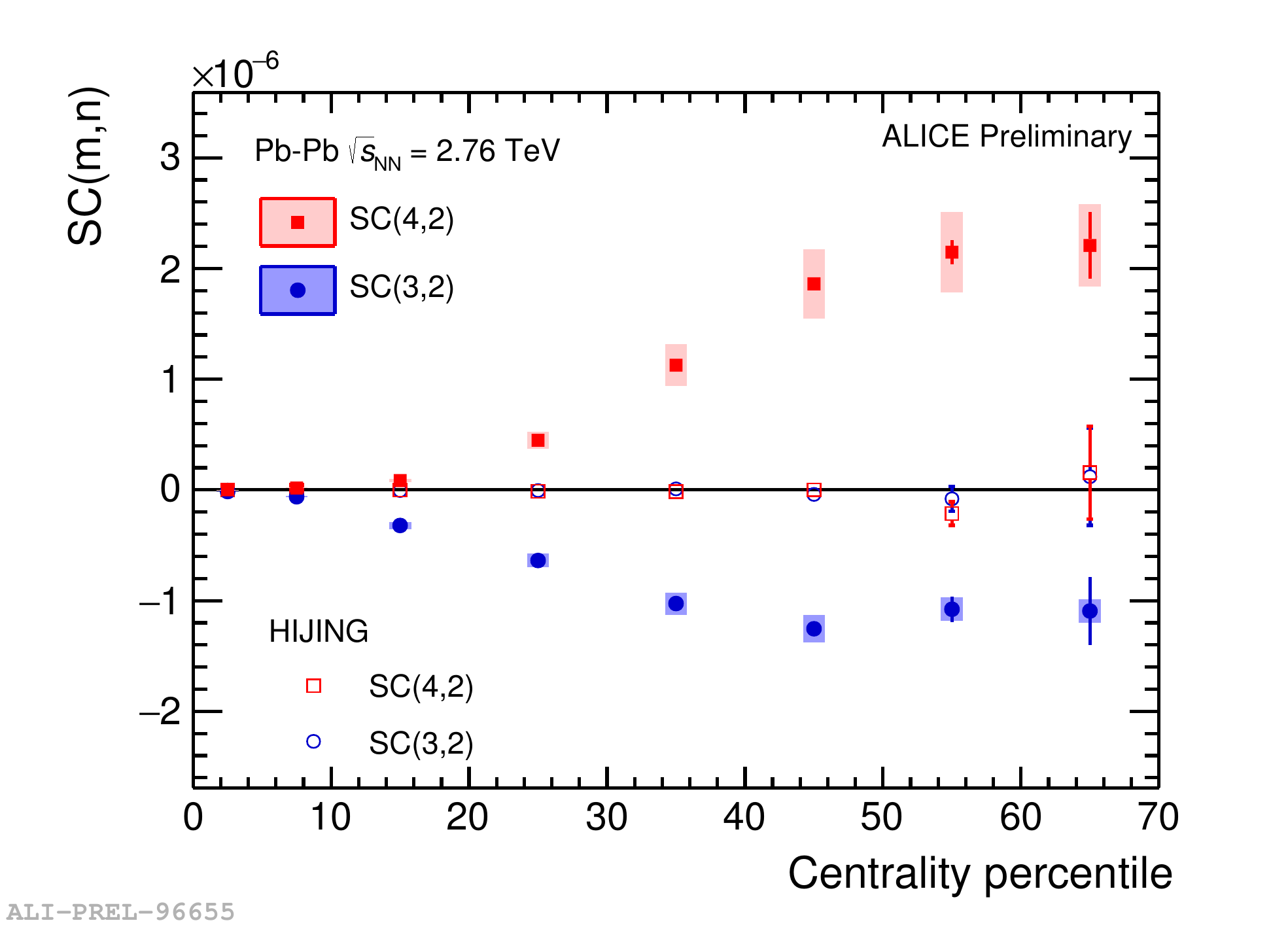}
\label{fig:figa41}}
\subfigure{\includegraphics[width=0.485\hsize]{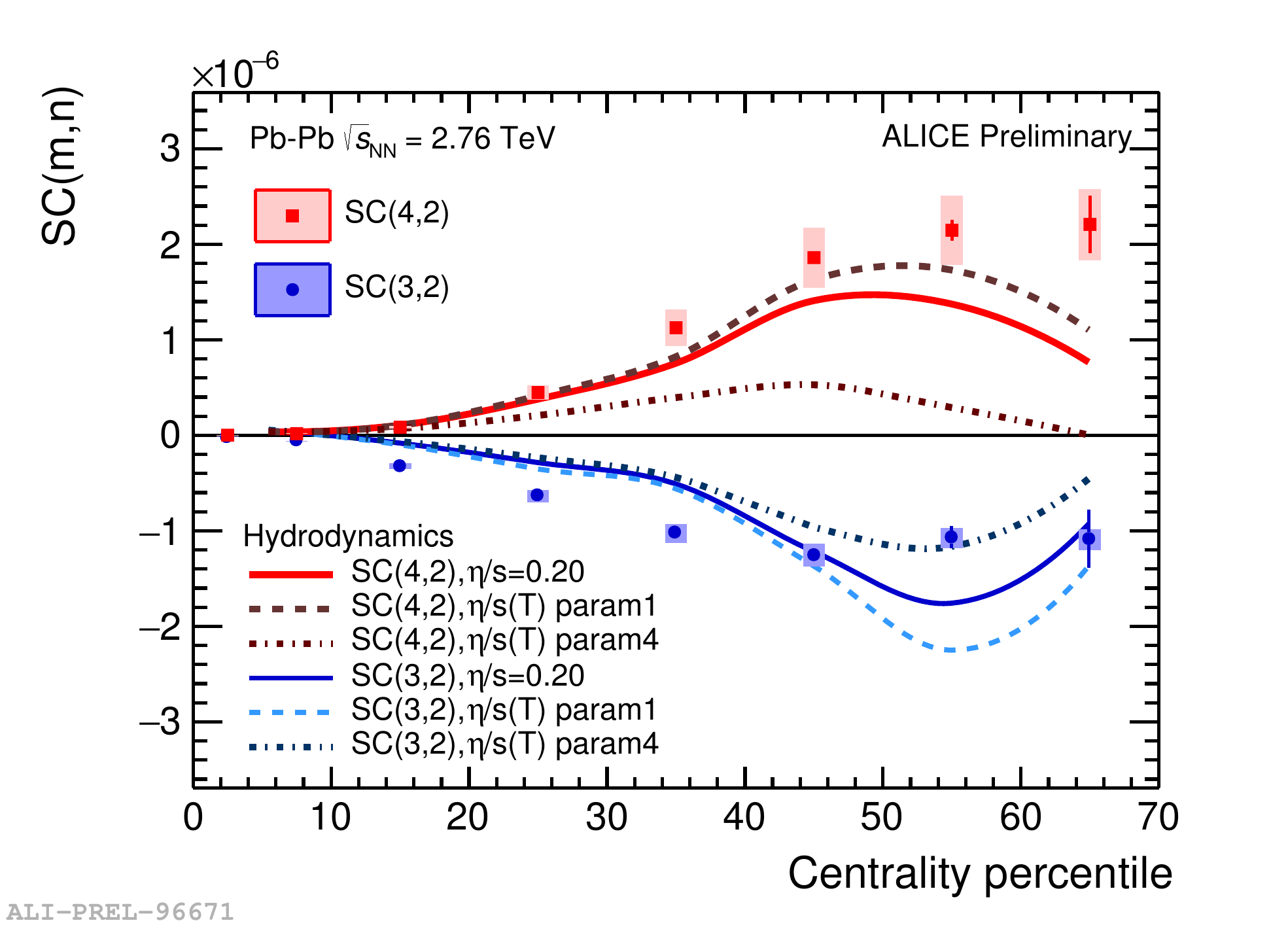}
\label{fig:figa42}}
\end{subfigmatrix}
\caption{(Color online) SC(4,2) and SC(3,2) as a function of centrality and comparison with HIJING with non-flow contributions (left) and hydrodynamical model calculations with different parametrization of $\eta/S(\rm T)$.}
\label{fig:fig4}
\end{figure}

The first measurements of the correlations between different flow harmonics have been performed by using 
symmetric 2-harmonics 4 particles cumulants, which 
is defined as ${\rm SC(m,n)} = $$<$$<$$\cos(m\psi_1+n\psi_2-m\psi_3-n\psi_4)$$>$$>$$_{\rm c}$ = 
$<$$v_{n}^2v_{m}^2$$>$-$<$$v_{n}^2$$>$
$<$$v_{m}^2$$>$, where $\psi$ is the azimuthal angle of particles.
Figure~\ref{fig:fig4} shows SC as a function of collision centrality, where boxes and circles correspond to 
${\rm SC(4,2)}$ and ${\rm SC(3,2)}$, respectively~\cite{bib:CC3}. Positive (negative) correlations are seen between 
$v_{2}$ and $v_{4}$ ($v_{3}$) and 
the correlations are not described by the HIJING generator (non-flow effects). 
These ${\rm SC(m,n)}$ values are sensitive to shear viscosity/entropy ratio ($\eta/S(T)$) 
of the created medium, 
as is shown in the right panel of Fig.~\ref{fig:fig4}, 
where 4 sets of $\eta/S(T)$ parametrization are used for the comparison. 
While centrality dependence of charged particle $v_{n}\{2\}$ is described by this parametrization~\cite{bib:SC}, 
SC measurements can provide stronger constrains on $\eta/S(T)$ than standard $v_{n}$ measurements. 

\subsection{Hard Probes}
Heavy flavors (i.e. charm and beauty) and jets are important probes for studying the properties of the medium 
since they are produced in the initial hard scattering and interact with the medium over the 
full evolution. ALICE released new results on the nuclear modification factor ($R_{AA}$) 
of $D$ and $D_{s}$ mesons, $R_{AA}$ and $v_{2}$ of forward muons from heavy-flavor semi-leptonic decays, 
charged jet $v_{2}$, 
and modifications of jet-core shape inside near side jets~\cite{bib:hf, bib:B2J, bib:jetb2, bib:jetm}.

\subsubsection{Heavy Flavors} 
\begin{figure}[htbp]
\begin{subfigmatrix}{2}
\subfigure{\includegraphics[width=0.45\hsize]{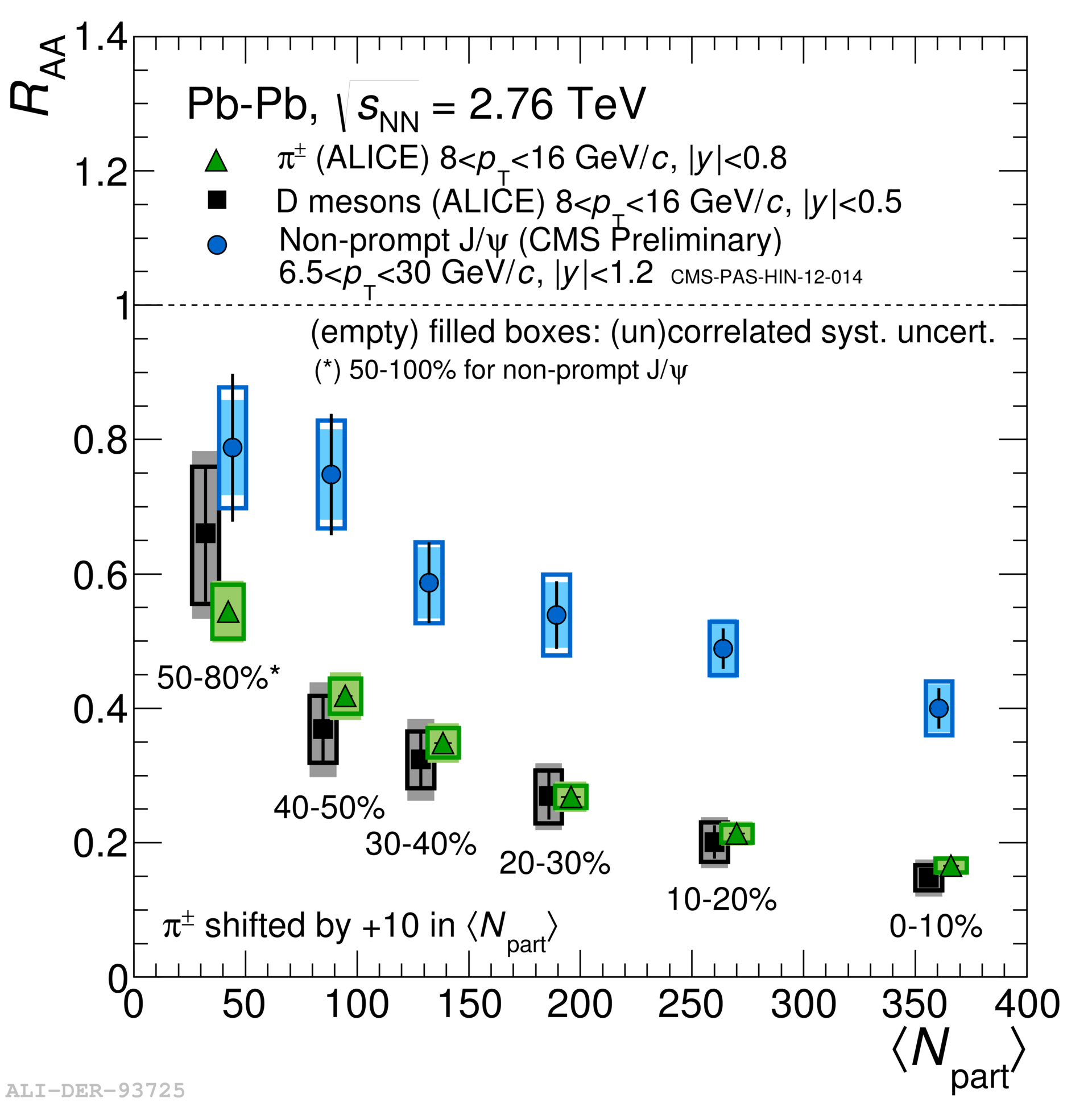}
\label{fig:figa51}}
\subfigure{\includegraphics[width=0.45\hsize]{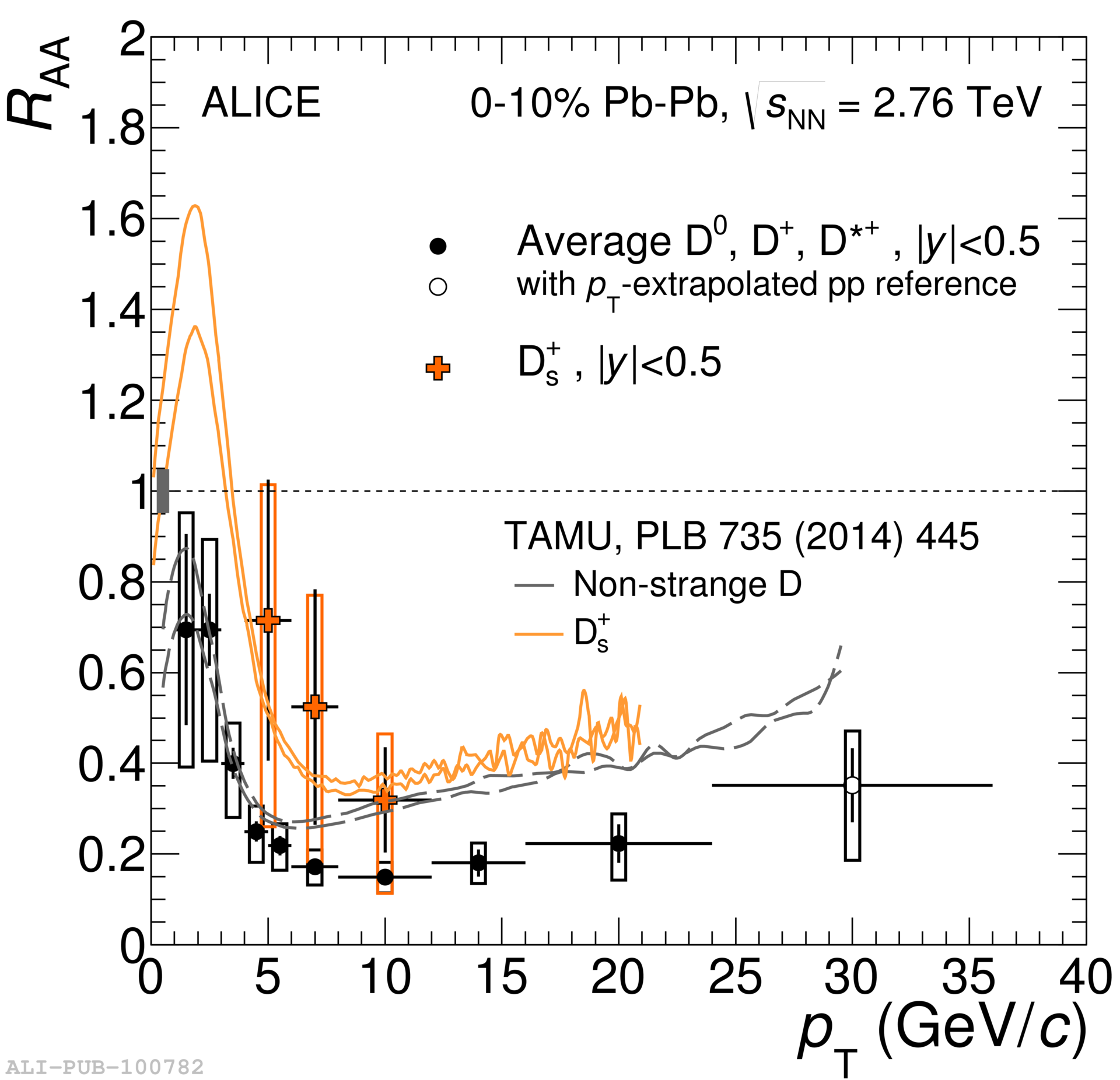}
\label{fig:figa52}}
\end{subfigmatrix}
\caption{(Color online) Left: $R_{AA}$ as a function of the number of participants ($N_{\rm part}$) for $\pi^{\pm}$, $D$, 
and non-prompt $J/\psi$ at similar $p_{\rm T}$ range. Right: $R_{AA}$ as a function of $p_{\rm T}$ for $D$ 
and $D_{s}$ mesons in 0-10\% centrality.}
\label{fig:fig5}
\end{figure}

Left of Fig.~\ref{fig:fig5} shows $R_{AA}$ for $\pi^{\pm}$, $D$, and non-prompt $J/\psi$ measured in 
similar $p_{\rm T}$ ranges (8$\le p_{\rm T} \le$ 16 GeV/$c$ for $\pi^{\pm}$ and $D$ 
and 6.5$\le p_{\rm T} \le$ 30 GeV/$c$ for non-prompt $J/\psi$) 
as a function of the number of participants. 
The $D$ meson $R_{AA}$ is much smaller than non-prompt $J/\psi$ $R_{AA}$ and 
is compatible with that of $\pi^{\pm}$ within uncertainties. 
These two observations are simultaneously described by models accounting for 
mass-dependent energy loss 
($\Delta E_{g} \ge \Delta E_{u,d,s} \ge \Delta E_{c} \ge \Delta E_{b} $), different shape of the parton $p_{\rm T}$
spectra and different parton fragmentation functions~\cite{bib:hf_eloss}.
Right of Fig.~\ref{fig:fig5} shows $R_{AA}$ as a function of $p_{\rm T}$ for $D$ 
and $D_{s}$ mesons in 0-10\% centrality. A hint of less suppression of $D_{s}$ mesons for $p_{\rm T} \le 8$~GeV/$c$ 
is visible, which is attributed to the strangeness enhancement in the medium and dominance of 
coalescence process for $D_{s}$ meson production.

\subsubsection{Jet-core shapes}
\begin{figure}[htbp]
\begin{subfigmatrix}{2}
\subfigure{\includegraphics[width=0.485\hsize]{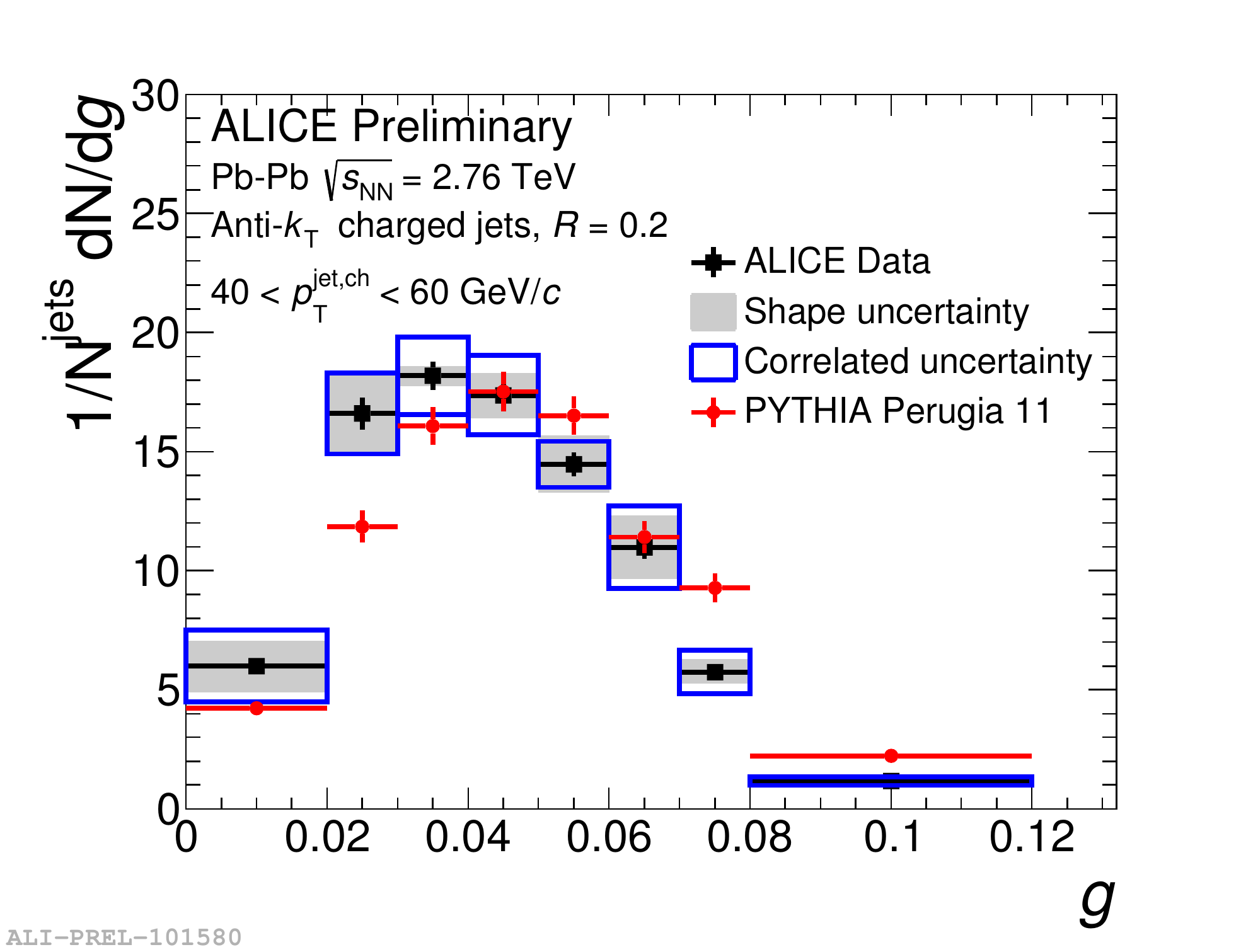}
\label{fig:figa61}}
\subfigure{\includegraphics[width=0.485\hsize]{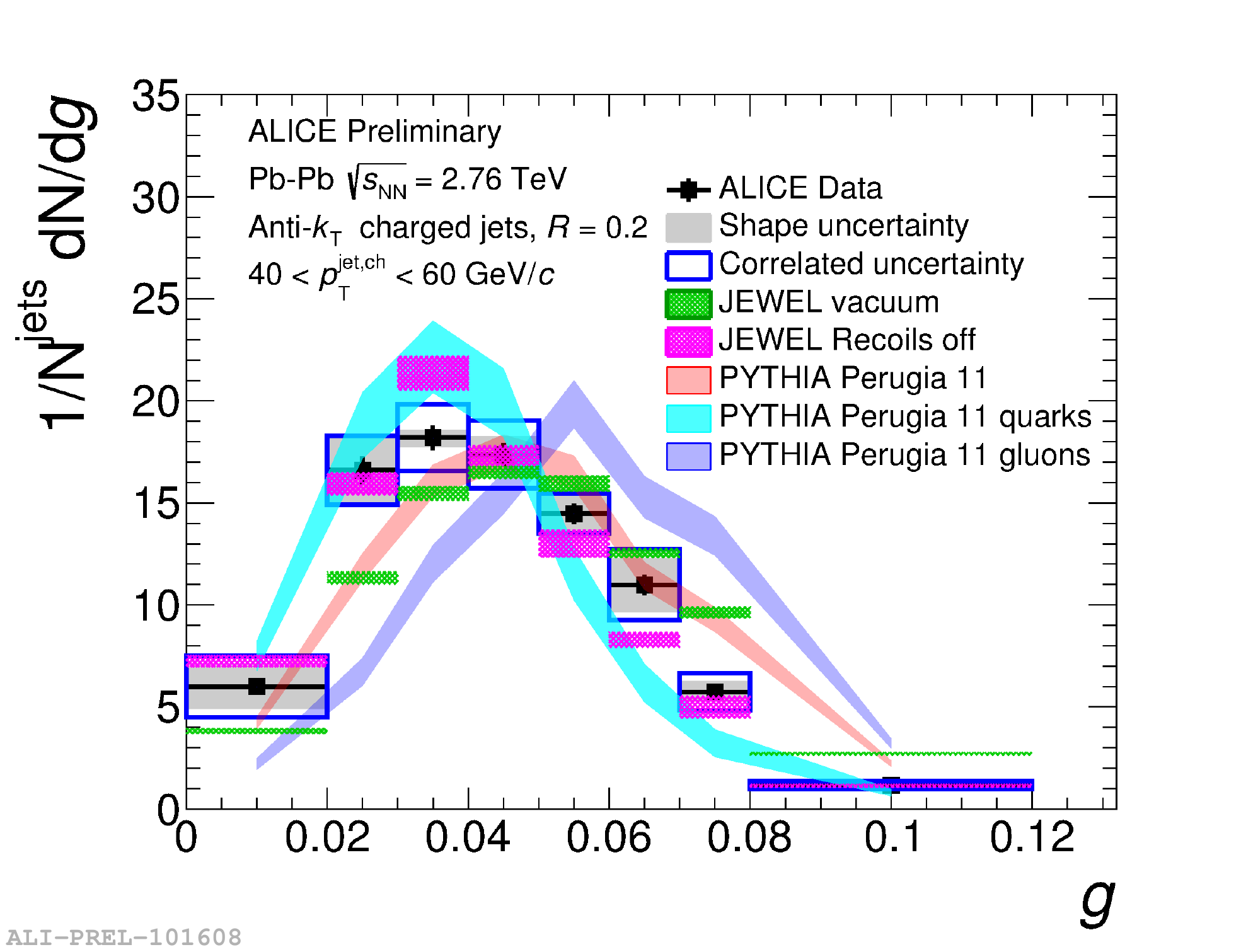}
\label{fig:figa62}}
\end{subfigmatrix}
\caption{(Color online) Left: Radial moment distributions of jet-core within $R = 0.2$ in Pb-Pb collisions 
and from PYTHIA Perugia 11. Right: Radial moment distributions of jet-core and comparison with 
JEWEL calculations with and without quenching and shapes from quark and gluon-jets calculated by PYTHIA Perugia 11.}
\label{fig:fig6}
\end{figure}

Three new observables are introduced and studied by ALICE to characterize the jet-core shapes:
radial momentum ($g = \Sigma p_{\rm T}^i |\Delta R_{i}|/{p_{\rm T}}_{\rm jet}$), dispersion in $p_{\rm T}$, and difference in $p_{\rm T}$ between leading and sub-leading jet constituents. 
Left of Fig.~\ref{fig:fig6} shows the radial moment distribution in 0-10\% top central collisions and the 
comparisons with PYTHIA Perugia 2011. 
Right of Fig.~\ref{fig:fig6} shows the comparison with JEWEL calculations with and without quenching and 
the quark and gluon jet core shapes estimated by PYTHIA Perugia 2011.
The radial moment distribution is shifted to lower values in Pb-Pb collisions when compared 
with pp collisions, 
which means that Pb-Pb jets are more collimated than pp jets.
This collimation is described by JEWEL model calculations 
indicating that the measured jet quenching and jet shape favor quark jets rather than gluon jets. 
This result suggests that gluon jets are more quenched than the quark jets resulting in a change 
of the quark/gluon jet relative composition in Pb-Pb collisions. 

\subsection{Low $p_{\rm T}$ Photons}  
ALICE has measured the direct photon spectra in Pb-Pb collisions for three centrality classes by using the 
photon conversion method and the PHOS calorimeter~\cite{bib:alice_photon}.
A 2.6 $\sigma$ excess of low $p_{\rm T}$ photons above pQCD calculations is observed in central 0-20\% collisions. 
The inverse slope parameter is extracted for 0-20\% centrality and found to be 304 $\pm$ 11$^{\rm stat}$ $\pm$ 40$^{syst}$ MeV, 30\% larger than the slope measured at RHIC.

\subsection{Longitudinal Asymmetry}  
\begin{figure}[htbp]
\begin{subfigmatrix}{2}
\subfigure{\includegraphics[width=0.36\hsize]{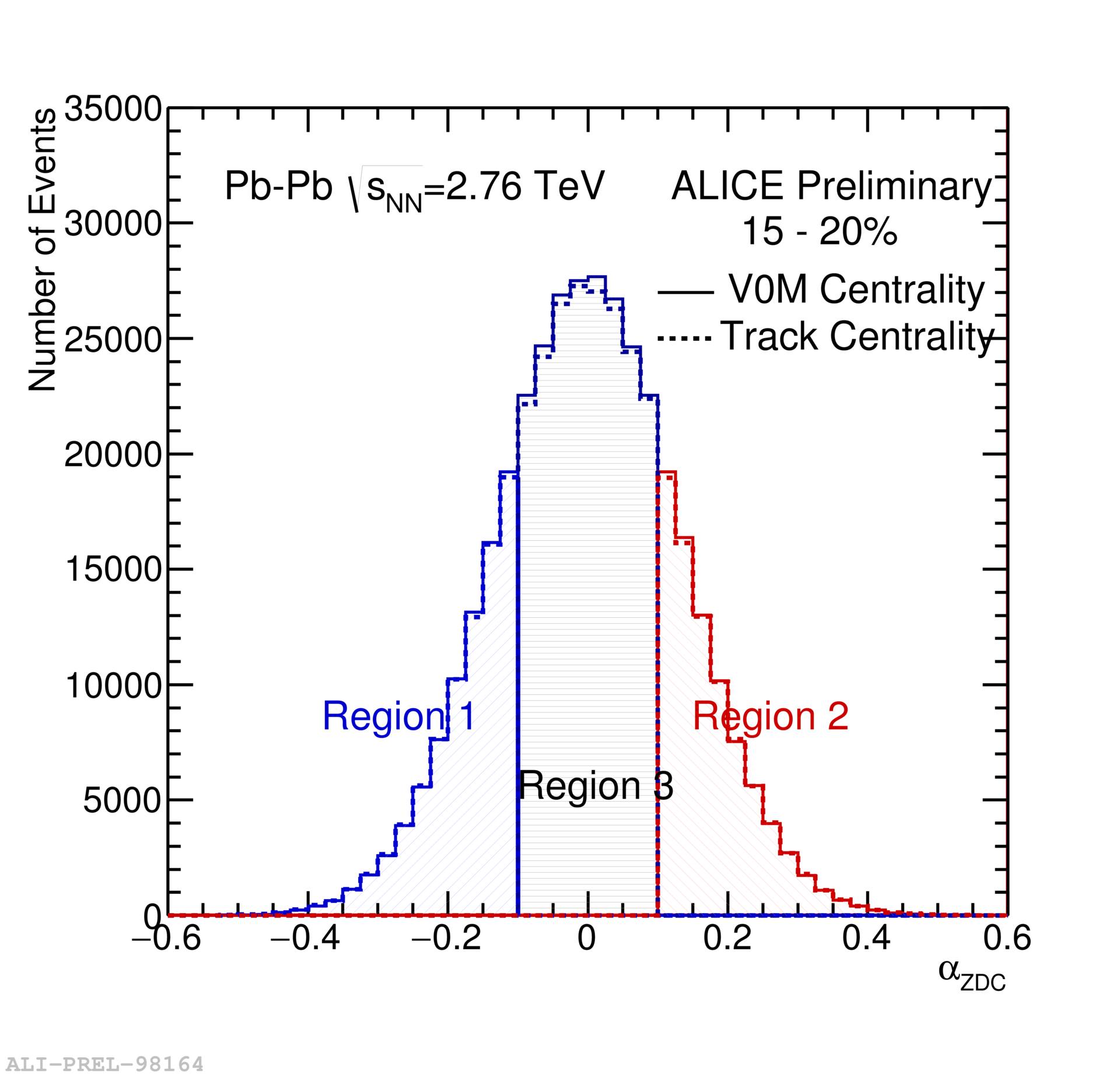}
\label{fig:figa_gl1}}
\subfigure{\includegraphics[width=0.62\hsize]{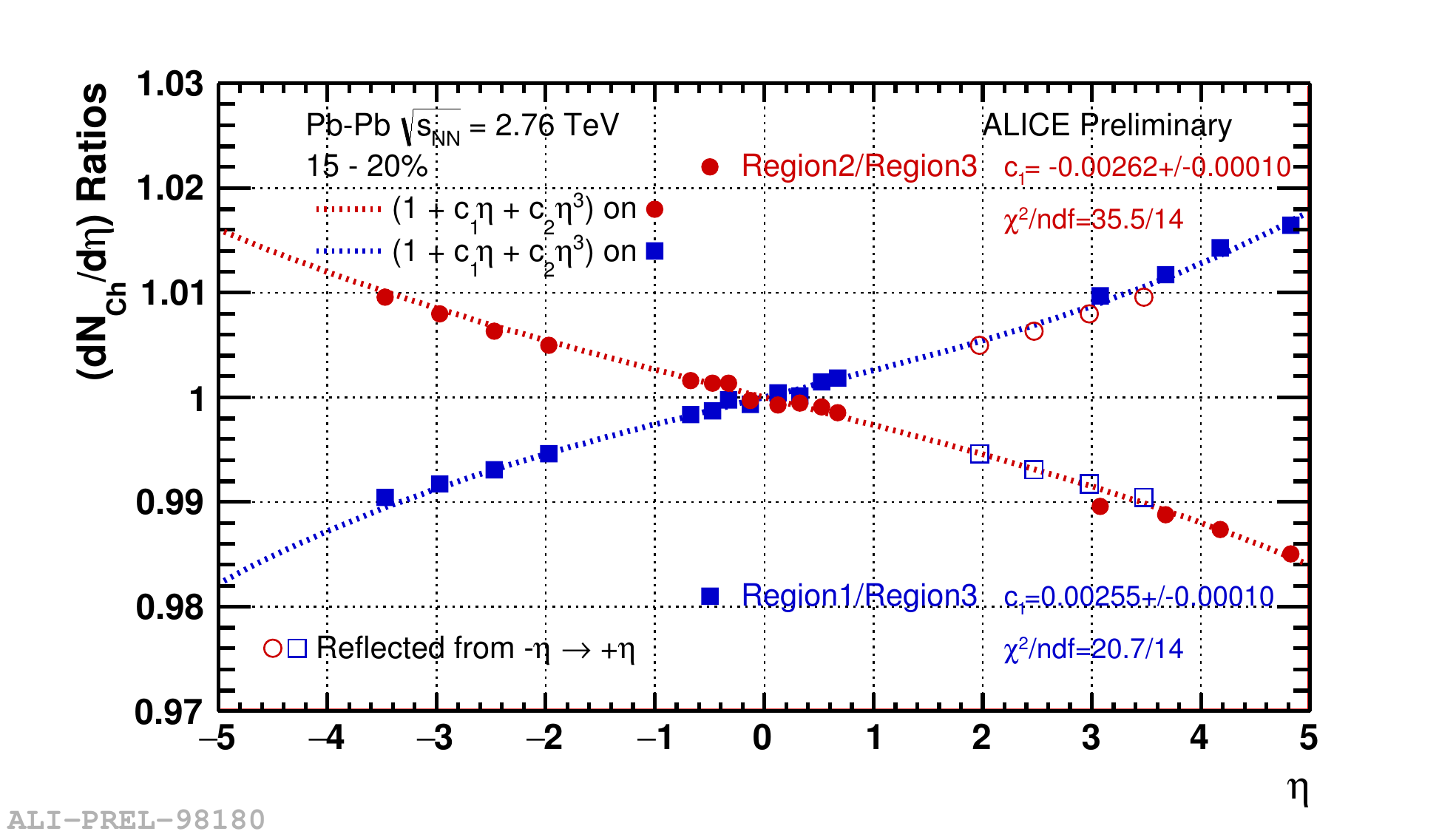}
\label{fig:figa_gl2}}
\end{subfigmatrix}
\caption{(Color online) Left: Measured energy asymmetry between two ZDCs in ALICE. Right: Ratio of $dN/d\eta$ between 
different measured energy asymmetric events. (circle: region2/region3, open:region1/region3)}
\label{fig:fig8}
\end{figure}

When two nuclei collide, the number of
participants for each one can be different due to the fluctuations of the nucleon density. 
These event-by-event fluctuations are estimated by measuring the asymmetry 
in the energy deposition in the two neutron ZDCs. 
The effect of these fluctuations on the $dN/d\eta$ distributions of charged particles has been studied
for the first time. Left of Fig.~\ref{fig:fig8} shows the distribution of the asymmetry 
in the energy deposition in the two ZDCs for 15-20\% centrality class~\cite{bib:alice_asym}.
Events have been further classified based on the measured energy asymmetry,
where Region-1, Region-2, and Region-3 covers $\alpha_{\rm ZDC} = (E_{\rm ZNA} - E_{\rm ZNC})/(E_{\rm ZNA} + E_{\rm ZNC}) < -0.1$, $\alpha_{\rm ZDC} > 0.1$, 
and $|\alpha_{\rm ZDC}|<0.1$, respectively.
The ratio of the measured $dN/d\eta$ distributions for Region-1 to Region-3 (box) and
Region-2 to Region-3 (circle) is shown in the right of Fig.~\ref{fig:fig8}.
One can clearly see a shift of the measured $dN/d\eta$ distributions for the events in Region-1 
and Region-2~\cite{bib:alice_asym}. Solid lines are the results of a fit to the ratios 
for both Region-1/Region-3 and Region-2/Region-3 with a cubic function. From the fitted coefficient, 
the values of the rapidity-shift is obtained and is found to be consistent 
with the estimations from a Glauber model calculation~\cite{bib:alice_asym}.


\section{Highlights from pp and p-Pb collisions}   
Recent intriguing results in heavy-ion physics raised questions on the origin of the non-trivial effects 
observed in high-multiplicity 
pp and p-Pb collisions such as the behavior of $<$$p_{\rm T}$$>$ and identified particle spectra as a function of 
multiplicity~\cite{bib:mul_pppb, bib:spectra_pppb}, long-range correlations~\cite{bib:lrc_pppb}, and 
mass ordering of $v_2$~\cite{bib:v2_pppb}. 
ALICE presented new exciting results, which include quarkonia and heavy-flavor production~\cite{bib:ppb_quarkonia, bib:ppb_hf} and $v_{2}$ of forward muons in p-Pb collisions~\cite{bib:ppb_fmu}, and strangeness production ($\Lambda$, $\Xi$, $\Omega$) 
in pp collisions~\cite{bib:pp_str}. 

\subsection{Cold and hot matter effects} 
Left and right of Fig.~\ref{fig:fig9} show the inclusive $J/\psi$~\cite{bib:ppb_jpsi} and $\psi$(2S) $Q_{\rm pPb}$ 
as a function of the number of binary collisions at backward (-4.46 $\le y_{\rm CMS} \le$ -2.96) and 
forward (2.03 $\le y_{\rm CMS} \le$ 3.53) rapidities, respectively~{\footnote{$Q_{\rm pPb}$ is the 
centrality-dependent nuclear modification factor defined as the yield ratios in A+A and pp collisions scaled by the 
number of binary collisions. This notation is used to warn for possible biases in the determination 
of the number of binary collisions.}
The centrality determination is based on the measured energy deposition in the two neutron ZDCs, which is the least biased estimator~\cite{bib:ppb_central}.
The $J/\psi$ production is compatible with the binary scaling at backward rapidity, 
while at forward rapidity the $J/\psi$ production is suppressed for all centrality classes. 
These trends are well-described by initial state effects such as gluon shadowing and initial state 
energy loss. The $\psi$(2S) suppression is visible in both rapidity intervals and increases with the centrality, 
where the final state interactions with a comoving medium and in the hadrons resonance gas 
is needed to explain the observed $\psi$(2S) suppression.

\begin{figure}[htbp]
\begin{subfigmatrix}{2}
\subfigure{\includegraphics[width=0.485\hsize]{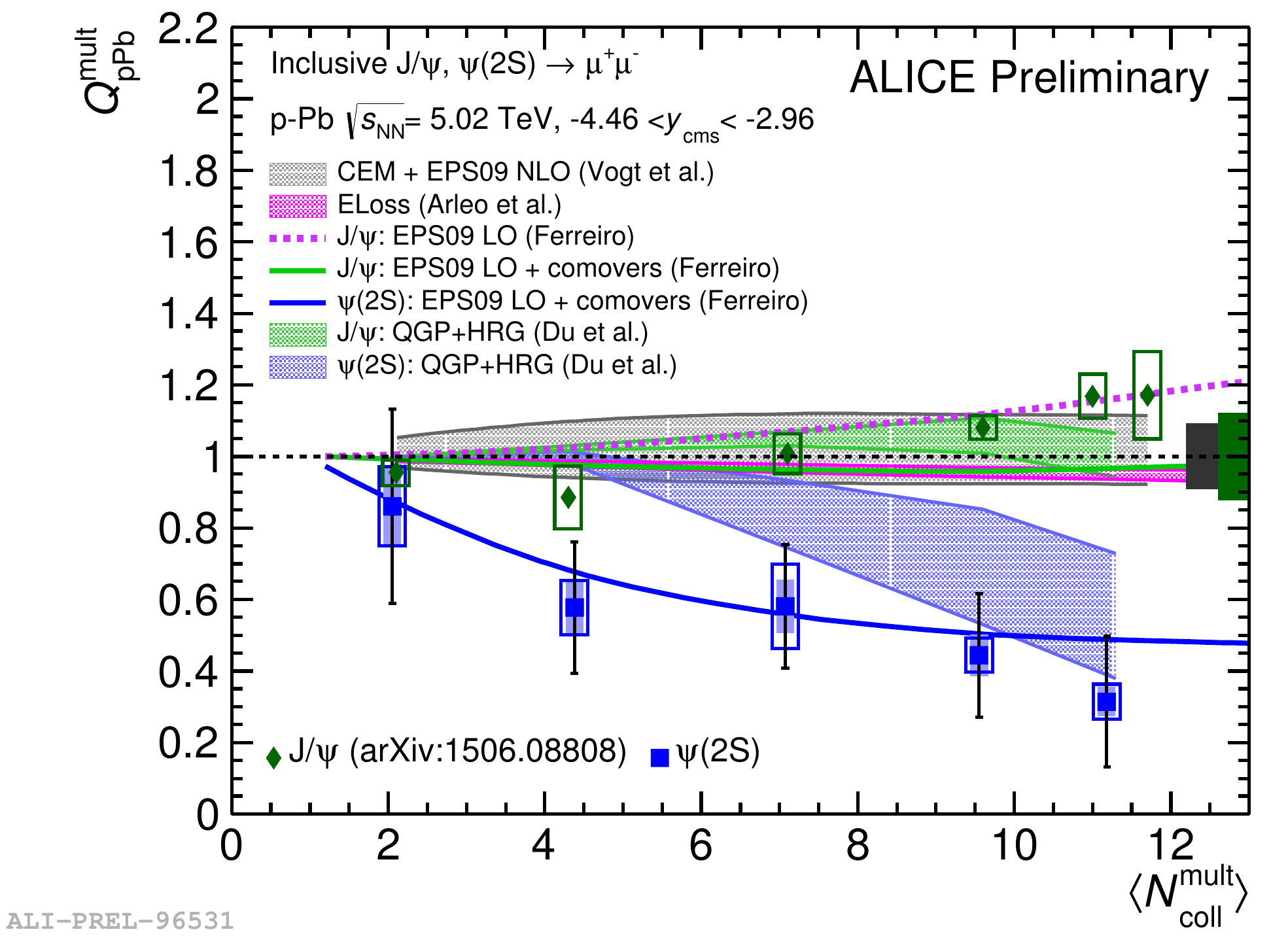}
\label{fig:figa_jpsi1}}
\subfigure{\includegraphics[width=0.485\hsize]{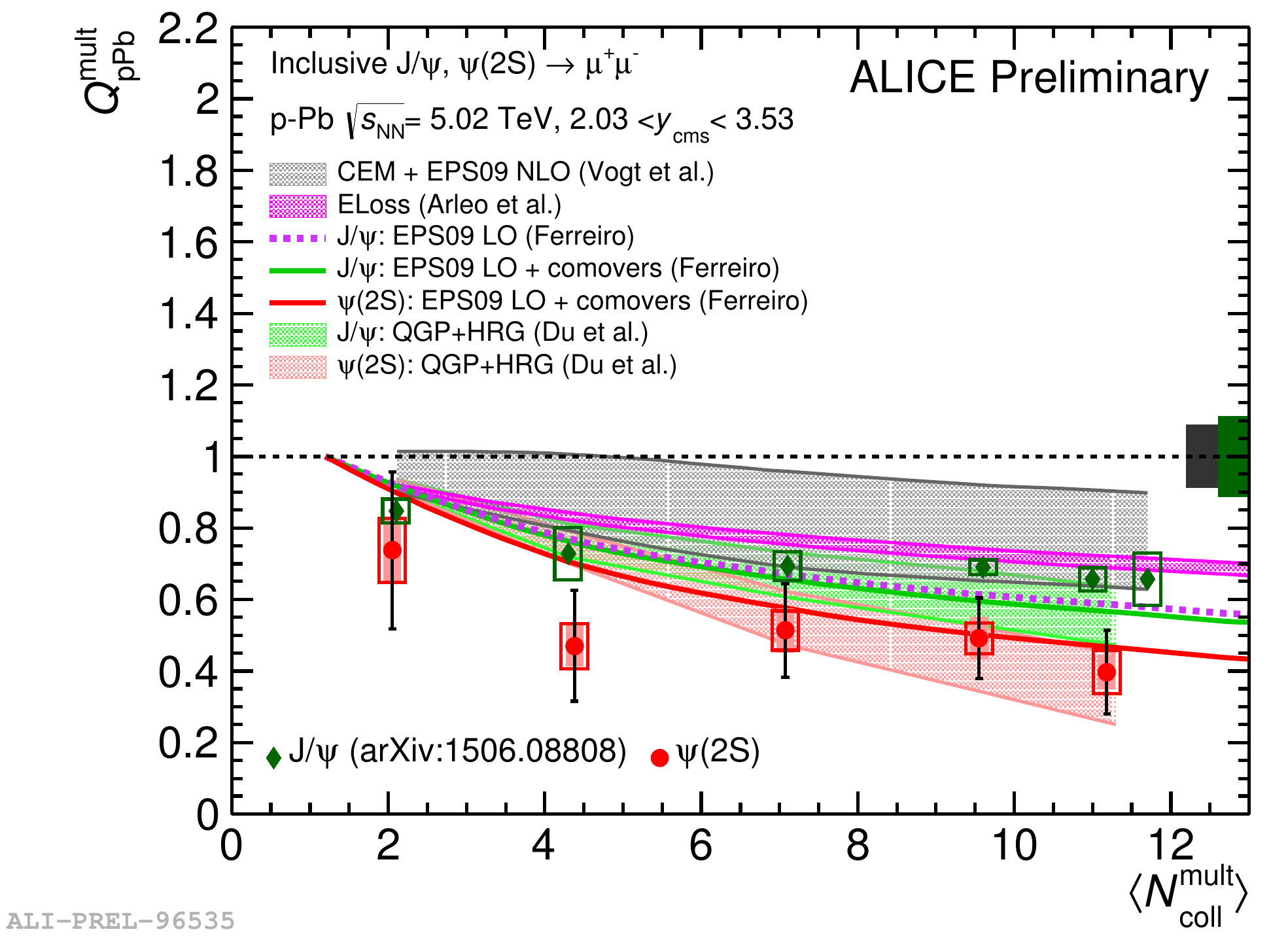}
\label{fig:figa_jpsi2}}
\end{subfigmatrix}
\caption{(Color online) $Q_{\rm pPb}$ as a function of forward rapidty (p-going direction) (left) and as a function 
of backward rapidity (Pb-going direction) (right) and comparison with models.}
\label{fig:fig9}
\end{figure}

\subsection{Forward muon $v_{2}$ in p-Pb collisions}
ALICE has measured long-range correlations at $\Delta \eta$ of up to 5 
by taking the correlations
between forward muons and midrapidity tracks or 
tracklets~\cite{bib:ppb_fmu}.
A double ridge structure is visible up to $\Delta \eta \sim$5 
and the extracted $p_{\rm T}$ differential $v_{2}$ of forward muons 
shows larger $v_{2}$ in Pb-going direction than in p-going 
direction, which is in accordance with AMPT calculations.
It is found that $v_{2}$ of forward muons above 2 GeV/$c$, where 60\% of muons 
are from heavy-flavor decays, is larger than zero.

\subsection{Heavy flavor production vs. event multiplicity in p-Pb collisions} 
Left and right of Fig.~\ref{fig:fig11} show the self-normalized yields of $D$ mesons as a function of 
relative multiplicity based on two different multiplicity estimators, 
where one uses the number of tracklets in SPD at $|\eta| \le$1 and the other uses 
multiplicities measured with V0A (Pb-going direction) covering 2.8 $\le \eta \le$ 5.1. 
For the SPD estimator, the results exhibit a faster-than-linear increase in the relative $D$ meson yields, 
while for the V0A estimator, the results show a roughly linear increase as a function of multiplicity
up to the interval of $N_{\rm V0A}$/$<$$N_{\rm V0A}$$>$ $\le$ 3.5.

\begin{figure}[htbp]
\begin{subfigmatrix}{2}
\subfigure{\includegraphics[width=0.45\hsize]{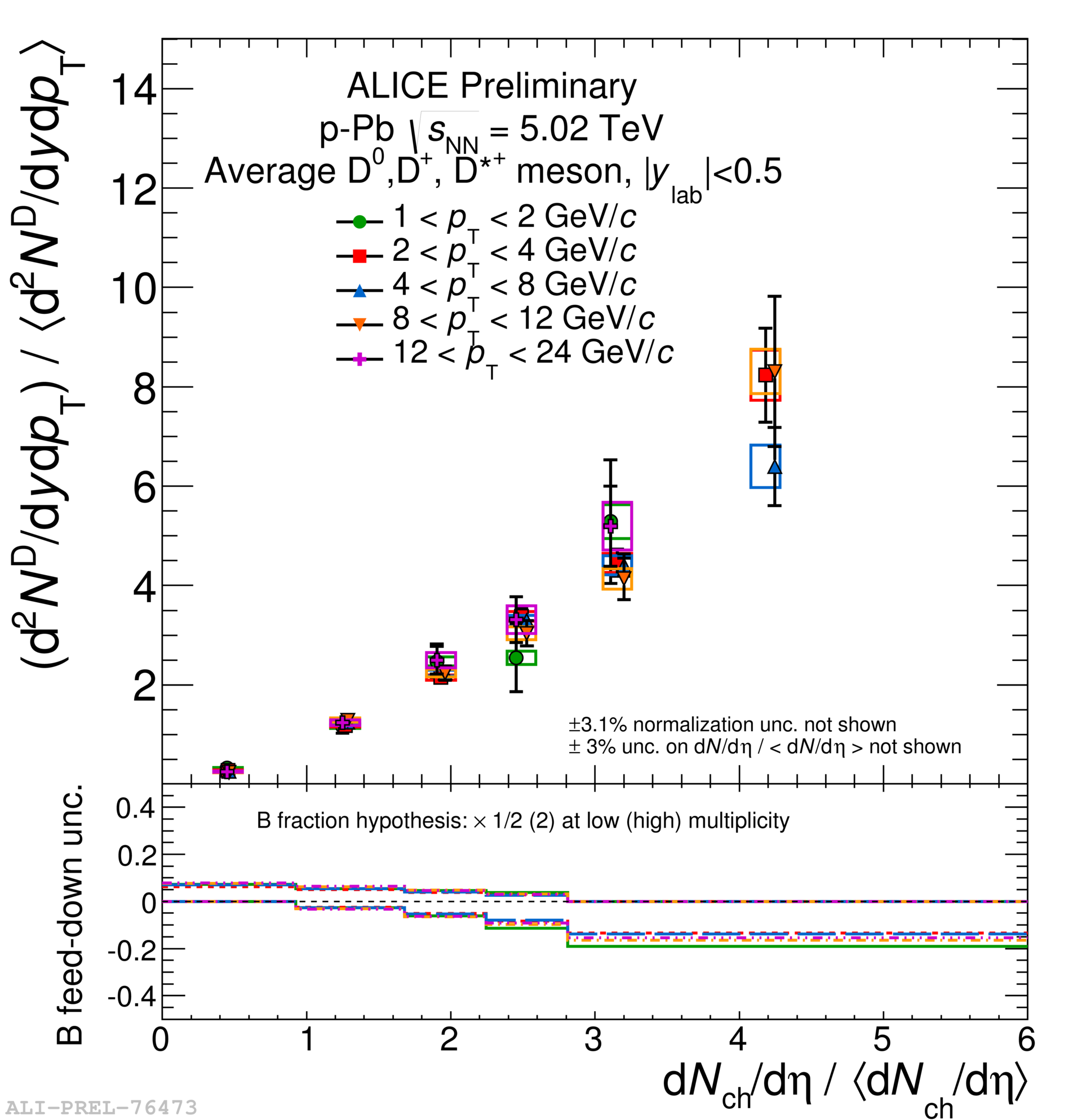}
\label{fig:figa71}}
\subfigure{\includegraphics[width=0.45\hsize]{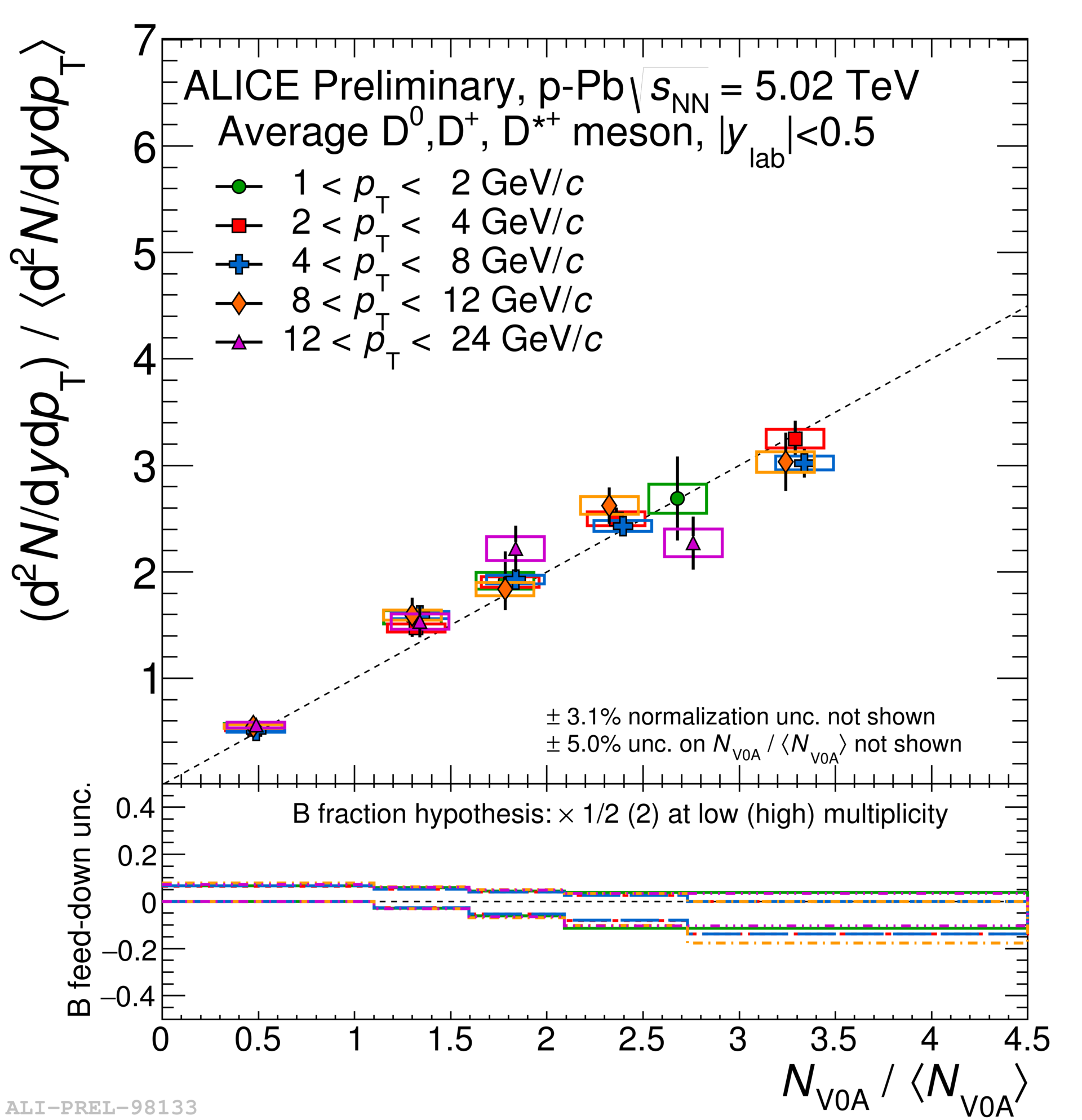}
\label{fig:figa¥72}}
\end{subfigmatrix}
\caption{(Color online) Self-normazlied yield as a function of relative multiplicity in $p-$Pb collisions. Relative multiplicity is estimated by SPD tracklets (left) and V0 detector at forward rapidity (right).}
\label{fig:fig11}
\end{figure}

\subsection{Strangeness production vs. event multiplicity in pp collisions}  
Figure~\ref{fig:fig12} shows the $p_{\rm T}$ differential $\Lambda$/$K^{0}_{s}$ ratio in the highest 
and lowest multiplicity classes for pp (left), p-Pb (middle), and Pb-Pb (right) collisions. 
The ratio depends on the event multiplicity and the magnitude of $\Lambda$/$K^{0}_{s}$ 
is qualitatively similar in 0-1\% pp, 60-80\% p-Pb, and 80-90\% Pb-Pb multiplicity classes, where 
the $<$$dN_{\rm ch}/d\eta$$>_{|\eta| \le 0.5}$ value is also comparable. 
Left of Fig.~\ref{fig:fig13} shows the yield ratios for proton, $\Lambda$, $\Xi$, and $\Omega$ to pion yields 
normalized to the values measured for pp INEL event class. 
Faster enhancement for $\Omega$$>$$\Xi$$>$$\Lambda$ is clearly visible, while the ratio for proton stays constant,   
and these trends are similar in pp and p-Pb collisions.
Right of Fig.~\ref{fig:fig13} shows the ratio to pions in pp, p-Pb, and Pb-Pb collisions 
normalized to the value in the highest multiplicity class in Pb-Pb collisions. 
Solid and dashed lines are the statistical model calculations and the decrease in the ratios is 
interpreted as due to the canonical suppression. 
\begin{figure}[htbp]
\begin{center}
\includegraphics[width=0.8\linewidth]{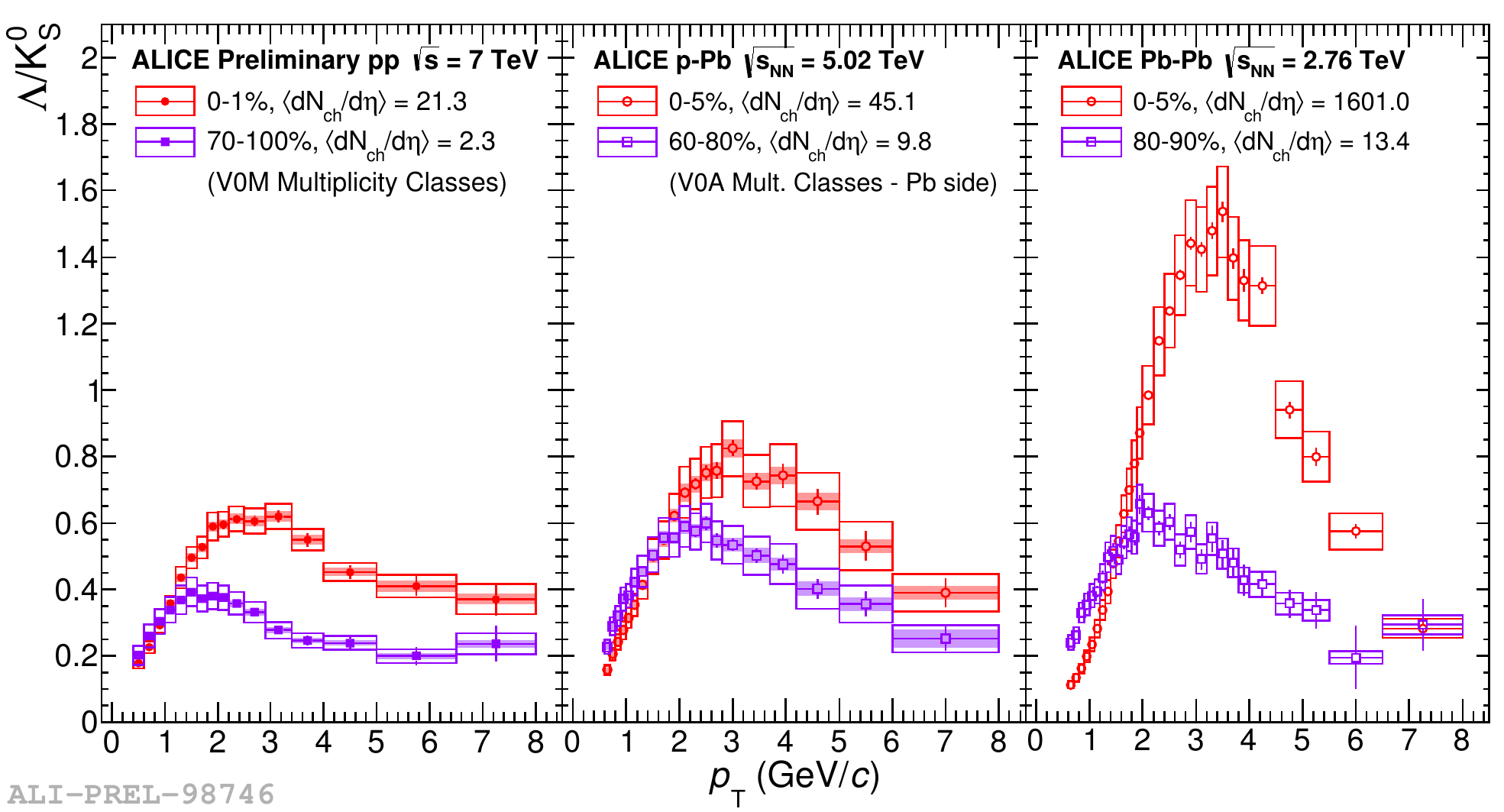}
\caption{(Color online) $(\Lambda+\bar{\Lambda})$/$K^{0}_{s}$ as a function of $p_{\rm T}$ in the highest and lowest multiplicity classes in pp (left), p-Pb (middle), and Pb-Pb collisions (right). }
\label{fig:fig12}
\end{center}
\end{figure}


\begin{figure}[htbp]
\begin{subfigmatrix}{2}
\subfigure{\includegraphics[width=0.40\hsize]{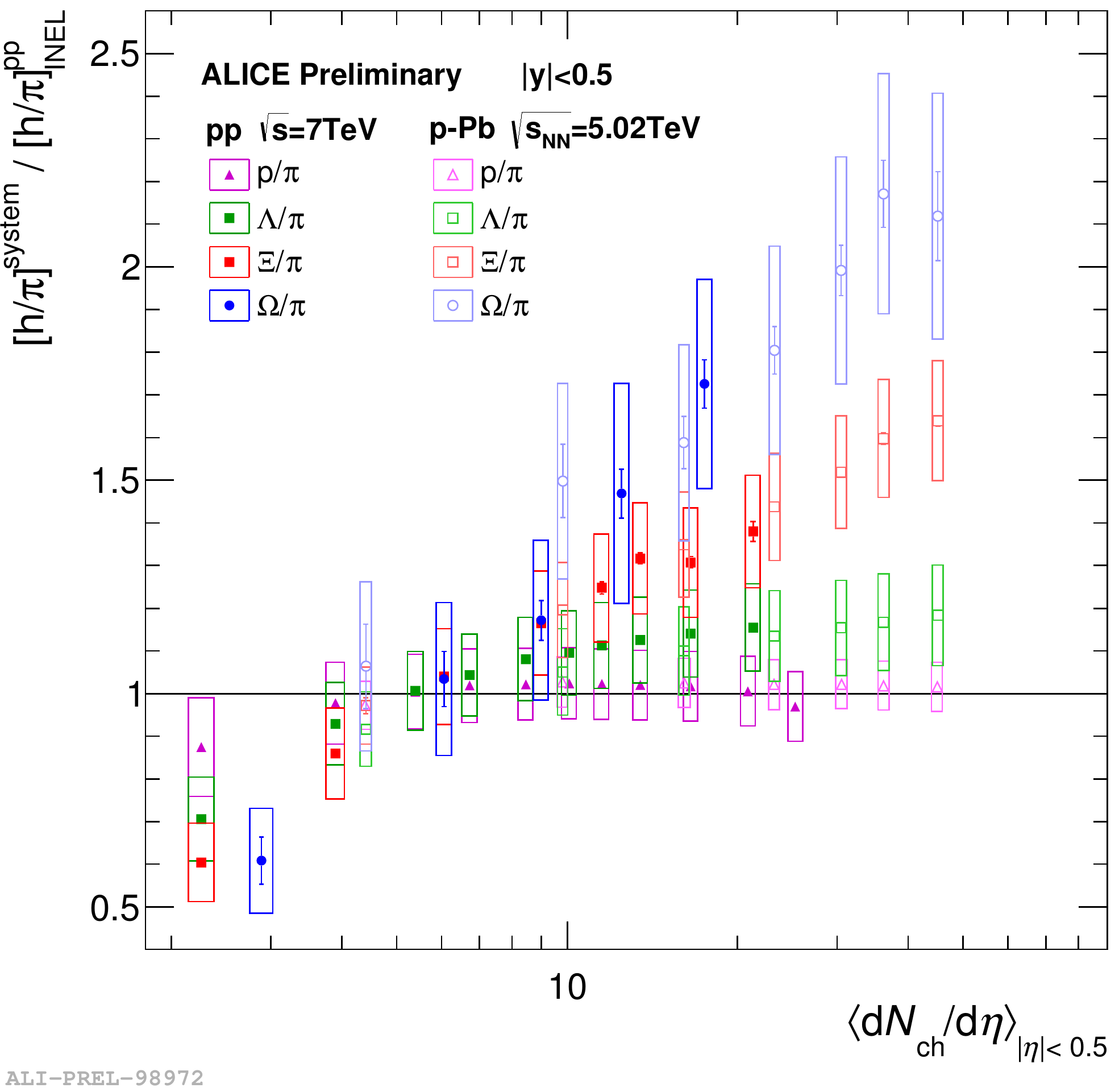}
\label{fig:figa_c1}}
\subfigure{\includegraphics[width=0.58\hsize]{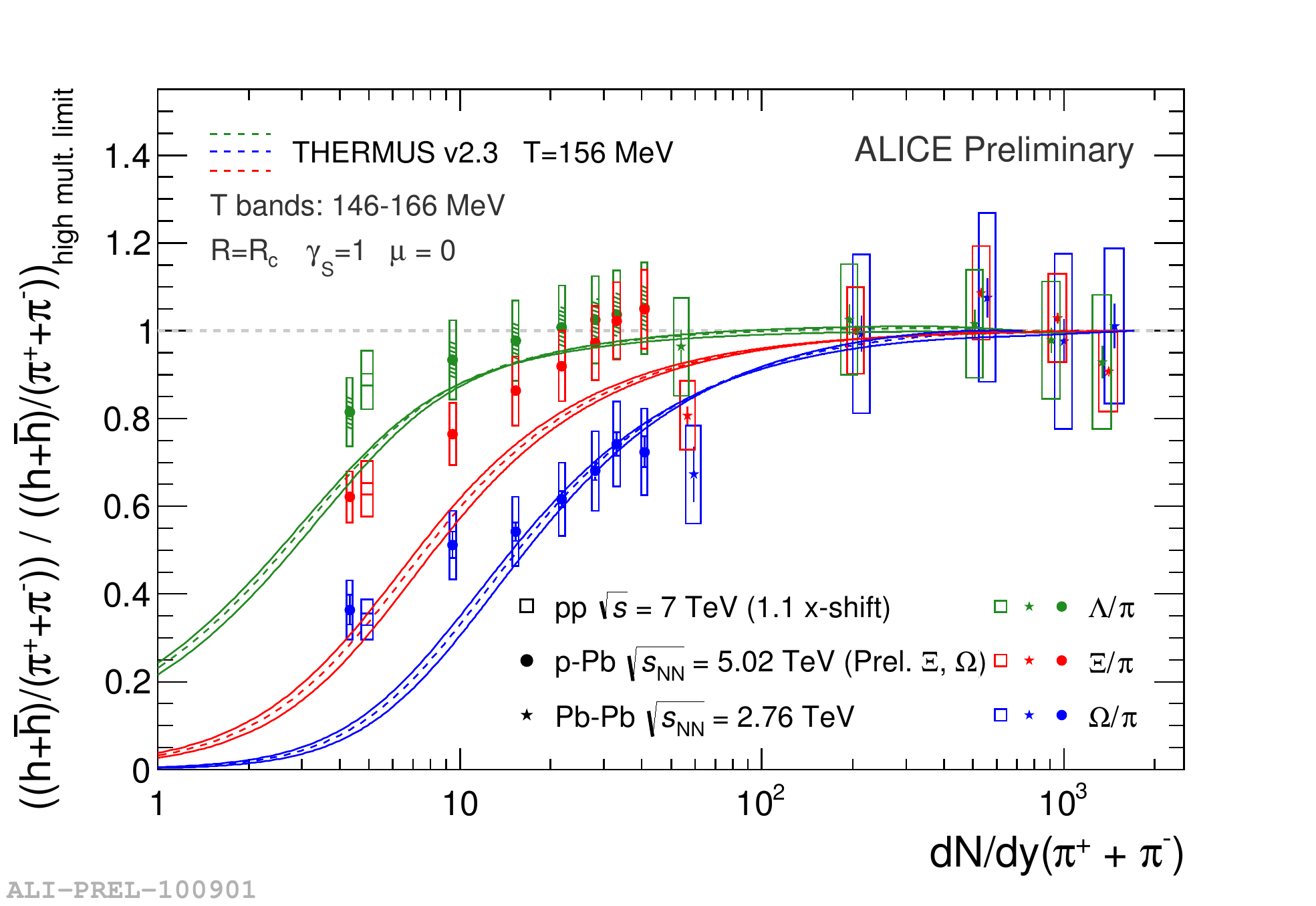}
\label{fig:figa_c2}}
\end{subfigmatrix}
\caption{(Color online) Left: Multi-strangeness ($\lambda$, $\Xi$, $\Omega$) and proton yield ratios to pions normalized to the values in pp 0-100\%. Right: Multi-strangeness yields normalized by the yields in the highest multiplicity class in Pb-Pb collisions.}
\label{fig:fig13}
\end{figure}

\section{ALICE in Run2 and Prospects for Run3}
During the Long Shutdown 1 of the LHC, the remaining five super modules of the TRD, 
eight modules of new dijet calorimeter (DCal), one module of PHOS, and 
new ALICE Diffractive detectors (AD) have been installed.
ALICE Central Trigger Systems have been upgraded to handle 100 trigger classes. 
The gas mixture of TPC has been changed from Ne(90):CO$_{2}$(10) to Ar(90):CO$_2$(10), to provide a more stable operation in high particle fluxes during heavy-ion running.
ALICE started taking data in pp collisions at $\sqrt{s} = 13$~TeV in June 2015. Charged particle multiplicities 
and $p_{\rm T}$ spectra for different multiplicity classes in pp INEL and INEL $>$ 0 events are reported 
in Ref.~\cite{bib:pp13TeV}.

The ALICE upgrades after Long Shutdown 2 (2019-2020) are focused on providing high precision 
measurements of dileptons, heavy flavors, quarkonia, jets 
and heavy nuclei with 100 times large statistics by taking minimum bias Pb-Pb collisions at 50 
kHz~\cite{bib:up0}. 
Many activities of the various ALICE upgrades including 
ITS, TPC, Muon Forward Tracker (MFT), trigger and readout electronics, and new combined online-offline project are on-going~\cite{bib:up1} 






\bibliographystyle{elsarticle-num}
\bibliography{<your-bib-database>}



\end{document}